\documentclass{aa} 
\usepackage{graphicx}
\begin{document}
   \title{No disk needed around HD 199143 B}
    \subtitle{}
\titlerunning{}
\authorrunning{Chauvin et al.}
\author{
        G. Chauvin\inst{1}
           T. Fusco\inst{2}
        A-M. Lagrange\inst{1}
        D. Mouillet\inst{3}
        J-L. Beuzit\inst{1}
        M. Thomson\inst{1}\inst{,4}
        J-C. Augereau\inst{5}
        F. Marchis\inst{6}
        C. Dumas\inst{7}
      \and
        P. Lowrance\inst{7}
}
%
    \institute{
                $^{1}$Laboratoire d'Astrophysique, Observatoire de 
Grenoble, 414, Rue de la piscine Domaine Universitaire Saint-Martin 
d'H\`eres, France\\
                $^{2}$ONERA, 29 Avenue de la Dividion Leclerc, 
Chatillon, France \\
                $^{3}$Laboratoire d'Astrophysique, Observatoire 
Midi-Pyr\'en\'enes, Tarbes, France\\
                $^{4}$Imperial College of Science, Technology and 
Medecine, Exhibition  Road, London SW7 2AZ, England\\
                $^{5}$DSM/DAPNIA/Service d'Astrophysique, CEA/Saclay, 91191 Gif-sur-Yvette,
France\\
                $^{6}$University of California Berkeley/Center for Adaptive Optics, 601 Campbell Hall, Berkeley, CA 94720, USA\\
                $^{7}$Jet Propulsion Laboratory, Pasadena, CA, USA\\
              }
\date{Received: 26 February 2002/ Accepted: 30 July 2002}
\abstract{
We present new, high angular resolution images of HD~199143 in the Capricornus association, obtained with the adaptive optics system ADONIS+SHARPII at the ESO 3.6m Telescope of La Silla Observatory. 
HD~199143 and its neighbour star HD~358623 (separation $\sim 5~\!'$ away) have previously been imaged with adaptive optics. For each star, a companion has been detected in the $J$ and $K$ bands at respective separations of $1.1~\!''$ and $2.2~\!''$ (Jayawardhana \& Brandeker 2001). Our new photometry of HD~199143~B suggests that it is a M2 star and that the presence of circumstellar dust proposed by Van den Ancker et al. (2000) is no longer necessary. We show that the 12~$\mu$m flux detected by IRAS previously interpreted as an IR excess, can be explained by the presence of the late-type companion.
}
\maketitle

%


\section{Introduction}

During the last few years, the nearby young stellar associations have become important targets for studying disk evolution and planet formation (Jayawardhana \& Greene 2001). Indeed, their proximity (typically closer than 100~pc) makes them ideal to observe the close circumstellar environment. Their youth (around 10~Myr) implies that planets may have already formed and that circumstellar disks may be optically thin and show evidence of spatial structure such as rings and gaps. The recent observations of the TW Hydrae Association (Kastner et al. 1997; Webb et al. 1999) or the Tucana Association  (Zuckerman et al. 2000; Zuckerman et al. 2001a) have revealed the existence of low-mass companions or circumstellar disks in these young and nearby open clusters. These detections are crucial to constrain the current models of proto-planetary systems formation and evolution. Searching more deeply into the close circumstellar environments of known associations and prospecting for new open clusters is clearly needed. The two physically bound stars HD~199143 and HD358623 have recently been presented as a new possible young association (the Capricornus Association), located at $\sim 48$~pc and aged  at $\sim 10$~Myr (Van den Ancker et al. 2000). They also have been associated recently to the Beta Pictoris Group by Zuckerman et al. (2001b). Spectroscopic observations have revealed the presence of ultraviolet and IR excess for HD~199143 which were attributed to the presence of an accretion disk around an hypothetical T~Tauri-like companion. 

To test these interpretations, high resolution imaging was needed. The first adaptive optics (AO) observations of the Capricornus association were carried out on 2001, May $31^{th}$ and June $1^{st}$. Two companions were resolved around both HD~199143 and HD~358623 (Jayawardhana \& Brandeker 2001). In the case of the binary system HD~199143, the ($J-K=1.37$) color observed for HD~199143~B was attributed to the  likely presence of a circumstellar disk. We present in this paper new results in $J$, $H$ and  $SK$ (short $K$) bands for this source, which lead to a different interpretation of the spectral type of HD~199143~B and the possible existence of a circumstellar disk around this companion. \vfill

\section{Observation and data reduction}

\begin{figure}[h]
\begin{center}
\vfill
\includegraphics[width = 4.25cm,angle = -90]{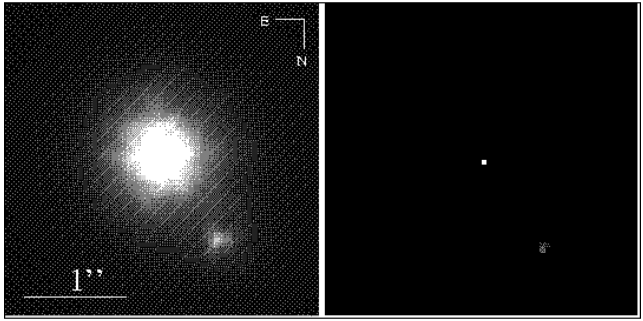}
\caption[]{\label{Fig1}{\small Left : ADONIS image in $J$ band of HD~199143 A and B (the pixel scale is 50 mas). Right : Resulting image obtained with the MISTRAL myopic deconvolution algorithm.}}
\label{fig:resdec}
\end{center}
\end{figure}

The observations were conducted at the La Silla Observatory on the 3.6m telescope  on 2001, June 2$^{nd}$ and 3$^{thd}$ and October 28$^{th}$ and 29$^{th}$. The AO system ADONIS combined with the SHARPII camera was used to observe HD~199143 in the near infrared broad bands $J$, $H$ and $SK$. A plate scale of $0.05~\!''$/pixel (field of view of $12.8~\!''\times12.8~\!''$) was chosen. In June, about 100 frames were taken with individual integration times of 0.4~s in $J$, 0.3~s in $H$ and 0.3~s in $SK$ with a seeing ranging from $0.85~\!''$ to $1.5~\!''$ in $V$ band. In October, we obtained 100 frames of 0.2~s in $H$ and $SK$ and 100 frames of 0.3~s in $J$ under good seeing conditions (stable at $0.9~\!''$).

The reference star HD~198420 was observed to estimate the PSF in each band during both periods. We also observed the stars S055D, S294D and S209D, taken from the HST photometric standards catalog, to estimate the PSF and obtain the near IR photometry of HD~199143~A and its companion HD~199143~B.      

The data were corrected from detector and sky background effects (bias, flat-fielding and badpixels correction). 
Then,  the myopic
deconvolution algorithm MISTRAL (Fusco et al. 1999, Conan et al. 2000) was used
to obtain the flux ratio and the separation of the system HD~199143. MISTRAL allows the
restoration of both the object and the Point
Spread Function (PSF). This technique is called ``myopic'' deconvolution because it accounts for the uncertainties and
variability of the PSF through the use of several images of a reference
star.  The \textit{a priori} knowledge of the object structure (multiple star) is also used. 
Fig.~\ref{fig:resdec} presents the result of the deconvolution on $J$ band images. 

\section{Results and comparison to previous data}

Based on the observations obtained in October 2001 (under the best photometric and seeing conditions), we derived  the photometry of HD~199143~A and its companion HD~199143~B in $J$, $H$ and $K$ bands. To obtain the photometry in $K$ band, we had to consider a correction factor to convert the flux from $SK$ to $K$ for both stars.  We took into account the spectral transmissions of both filters and the difference of spectral type between the science object (F8V star HD~199143~A and M2 star HD~199143~B) and an A0V star. In both cases, this factor was smaller than 0.004 magnitude and therefore neglected. The separation between the two components was estimated by MISTRAL at $1.074 \pm 0.004~\!''$ (consistent in $J$, $H$ and $K$ for observations of June and October 2001). The new photometric results are reported in Table~\ref{tab:phot}. The contrasts between HD~199143~A and B during the three AO observations are also compared in Table~\ref{tab:cont}. Jayawardhana \& Brandeker (2001) used the myopic deconvolution algorithm IDAC for their data set.
For HD~199143~A, our results are consistent with the values obtained by Jayawardhana \& Brandeker (2001)  respectively $J = 6.23 \pm 0.04$ and $K = 5.90 \pm 0.03$ if we consider our photometric uncertainties of 0.04 magnitude. 
These results are also consistent with the photometric values derived for unresolved observations of HD~199143 presented by Oudmaijer et al. (2001).

But, interestingly, the contrasts we measured between HD~199143 A and B at $J$ band are significantly different from those measured by Jayawardhana \& Brandeker (2001) as we will discuss in the next section. 
 \begin{table}[h]
      \caption[]{Photometry of HD~199143 A and B (October 2001).}
         \label{tab:phot}
         \centering
      \[
          \begin{array}{llll}
            \hline
    \hline
\noalign{\smallskip}
           \mathrm{Sources}             & \mathrm{J}             & \mathrm{H}             & \mathrm{K}  \\
                                        &\mathrm{(mag.)} &\mathrm{(mag.)} &\mathrm{(mag.)} \\
\noalign{\smallskip}
            \hline
\noalign{\smallskip}
            \mathrm{HD~199143~A+B}        & 6.19          & 5.90          & 5.82       \\
          \mathrm{HD~199143~A}            & 6.27  & 6.02          &5.95     \\
          \mathrm{HD~199143~B}            & 8.95  & 8.35          & 8.14    \\
\noalign{\smallskip}
 \hline
         \end{array}
      \]
 \end{table}
 \begin{table}[h]
     \caption[]{Comparaison of actual contrast observations between HD~199143 A and B.}
        \label{tab:cont}
      \[
         \begin{array}{lllll}
            \hline
    \hline          
 \noalign{\smallskip}
           \mathrm{Epoch}        & \Delta\mathrm{J}               & \Delta \mathrm{H}              & \Delta \mathrm{K}  & \mathrm{Dec.~Algo.}    \\
        \noalign{\smallskip}
            \hline
        \noalign{\smallskip}
           \mathrm{01/06~(a)}     & 3.25\pm0.04           & -             & 2.21\pm0.03  & \mathrm{IDAC}  \\
           \mathrm{02/06~(b)}              &2.67\pm0.05    & -     & 2.16\pm0.02         & \mathrm{MISTRAL}  \\
           \mathrm{29/10~(c)}           & 2.68\pm0.03   &2.33\pm0.01    & 2.19\pm0.02        & \mathrm{MISTRAL}    \\
\noalign{\smallskip}
  \hline
         \end{array}
      \]
\begin{list}{}{}
\item[$^{\mathrm{a}}$]  (Jayawardhana \& Brandeker 2001)
\item[$^{\mathrm{b}}$]  (this paper)
\item[$^{\mathrm{c}}$]  (this paper)
\end{list}
 \end{table}

\section{Discussion of contrast estimation}

In the case of HD~199143~B, our photometry presented in Table \ref{tab:phot} is different from the one presented by Jayawardhana \& Brandeker (2001) as we found a significant difference of contrast in $J$ band. This difference is supported by the results obtained in June and October 2001. Instrumental effects can be dismissed because the observations were taken with the same instrumental configuration (ADONIS+SHARPII) on the ESO 3.6m telescope. Uncertainties in our photometric measurements cannot be the source of this 0.57 magnitude difference since our K-band measurements are in perfect agreement between the two observation dates (02/06 and 29/10). Moreover, the $J$ and $K$ band contrast measurements that we obtained between HD~358623~A and HD~358623~B are consistent with the results of Jayawardhana \& Brandeker (2001) as presented in Table~\ref{tab:cont2}. In this case, the separation between the two components is estimated at $\sim2.2~\!''$ and is more favorable for the deconvolution.
We propose to investigate the two causes which may be responsible for this discrepancy: the variability of the stars and the two myopic deconvolution algorithms used IDAC and MISTRAL.

\begin{table}[h]
     \caption[]{Comparaison of actual contrast observations between HD~358623 A and B.}
        \label{tab:cont2}
      \[
         \begin{array}{llll}
            \hline
      \hline
            \noalign{\smallskip}
            \mathrm{Epoch} & \Delta \mathrm{J}   & \Delta \mathrm{K} & \mathrm{Dec.~Algo.} \\
        \noalign{\smallskip}
            \hline
        \noalign{\smallskip}
           \mathrm{01/06~(a)} & 1.81\pm0.08    & 1.67\pm0.06     & \mathrm{IDAC}  \\
           \mathrm{02/06~(b)} & 1.75\pm0.06     & 1.67\pm0.05         & \mathrm{MISTRAL}  \\
        \noalign{\smallskip}
            \hline
         \end{array}
      \]
\begin{list}{}{}
\item[$^{\mathrm{a}}$]  (Jayawardhana \& Brandeker 2001)
\item[$^{\mathrm{b}}$]  (this paper)
\end{list}
\end{table}

\subsection{Variability or flaring}

A first explanation might be the photometric variability of one or both stars in this sytem; this phenomenon is relatively frequent for young T~Tauri stars. However, Carpenter et al. (2001) have observed near-infrared photometric variability of Classical and Weak T Tauri stars toward the Orion A molecular cloud. The mean peak to peak amplitudes they detected were $\sim0$.2 magnitude, less than the present difference of 0.57 magnitude. In addition, even if the case of a flaring star is considered, a variability observed in $J$ band would also be expected in $K$ band which is not the case. Consequently, it is unlikely that this difference originates from stellar processes. \\

\subsection{Deconvolution algorithms}

\begin{figure}[hb]
\begin{center}
\includegraphics[width = 8.8cm,angle=-90]{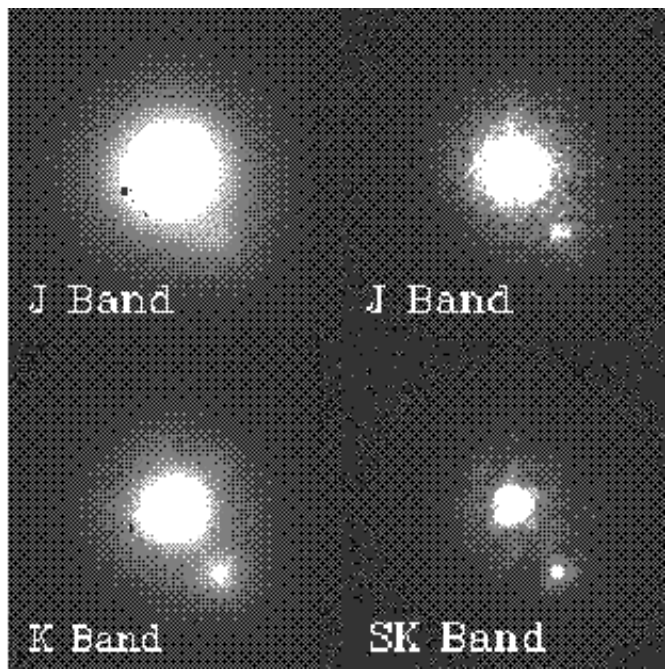}
\caption[]{\label{Fig2}{\small \textbf{Left :} ADONIS images in $J$ and $K$ band of HD~199143~AB obtained by Jayawardhana \& Brandeker (2001).  \textbf{Right :} ADONIS images we have obtained in $J$ and $SK$. The images are scaled to the same pixel scale (35~mas) and the same integration time (1~s).}}
 \label{fig:comp}
\end{center}
\end{figure}

The discrepancy of the $J$ band photometry is present between two distinct data sets deconvolved with two different deconvolution algorithms MISTRAL (that we used) and IDAC (used by Jayawardhana \& Brandeker (2001)). We first investigated the difference in the image quality of both observations as shown in Fig.~\ref{fig:comp}. Jayawardhana \& Brandeker (2001) kindly shared their data to conduct this analysis. The AO correction in their observations is low compared to our data (due probably to worse weather conditions). Our observations are then more favorable to perform myopic deconvolution and deduce the photometry of HD~199143~B (see Table~\ref{tab:phot}). If we perform deconvolution with IDAC on our observation, the results are consistent with MISTRAL deconvolution results (see Table~\ref{tab:deconv}).

In the unfavorable case of a low AO correction, we can still perform deconvolution but the regularisation on the PSF and the object needs to be done cautiously. The results obtained with MISTRAL and IDAC on Jayawardhana \& Brandeker (2001) observations are also presented in Table~\ref{tab:deconv} and show that this regularisation seems to be responsible for the discrepancy in $J$ band.

\begin{table}[h]
     \caption[]{MISTRAL and IDAC results compared for the observations obtained by Jayawardhana \& Brandeker (2001) in (01/06) and our observations in (29/10) in the case of HD 199143.}
        \label{tab:deconv}
      \[
         \begin{array}{llll}
            \hline
      \hline
            \noalign{\smallskip}
            \mathrm{Epoch} & \Delta \mathrm{J}   & \Delta \mathrm{K} & \mathrm{Dec.~Algo.} \\
        \noalign{\smallskip}
            \hline
        \noalign{\smallskip}
           \mathrm{01/06~(a)} & 3.25\pm0.04    & 2.21\pm0.03     & \mathrm{IDAC}  \\
           \mathrm{01/06~(a)} & 2.73\pm0.1     & 2.14\pm0.1         & \mathrm{MISTRAL}  \\
        \noalign{\smallskip}
            \hline
        \noalign{\smallskip}
           \mathrm{29/10~(c)} & 2.72\pm0.05      & 2.25\pm0.05   & \mathrm{IDAC}  \\
           \mathrm{29/10~(c)} & 2.68\pm0.03    & 2.19\pm0.02         & \mathrm{MISTRAL}  \\
  \noalign{\smallskip}
  \hline
         \end{array}
      \]
\begin{list}{}{}
\item[$^{\mathrm{a}}$] (Jayawardhana \& Brandeker 2001)
\item[$^{\mathrm{c}}$] (this paper)
\end{list}
\end{table}

\section{Astrophysical results}

\subsection{HD~199143~A}

\begin{figure}[htb]
\begin{center}
\vspace{0cm}
\hspace{0cm}\includegraphics[width = 7.7cm]{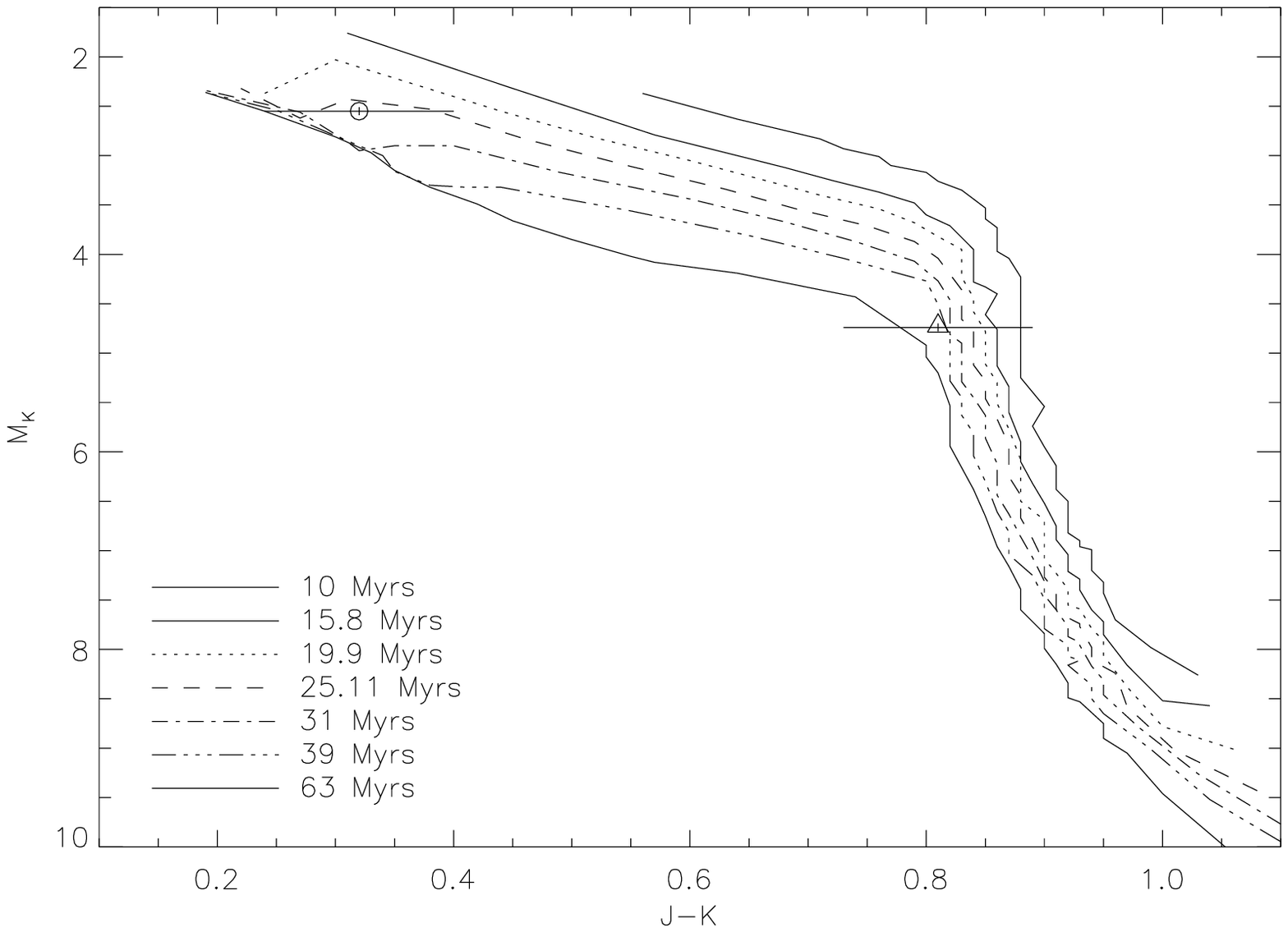}
\includegraphics[width = 7.7cm]{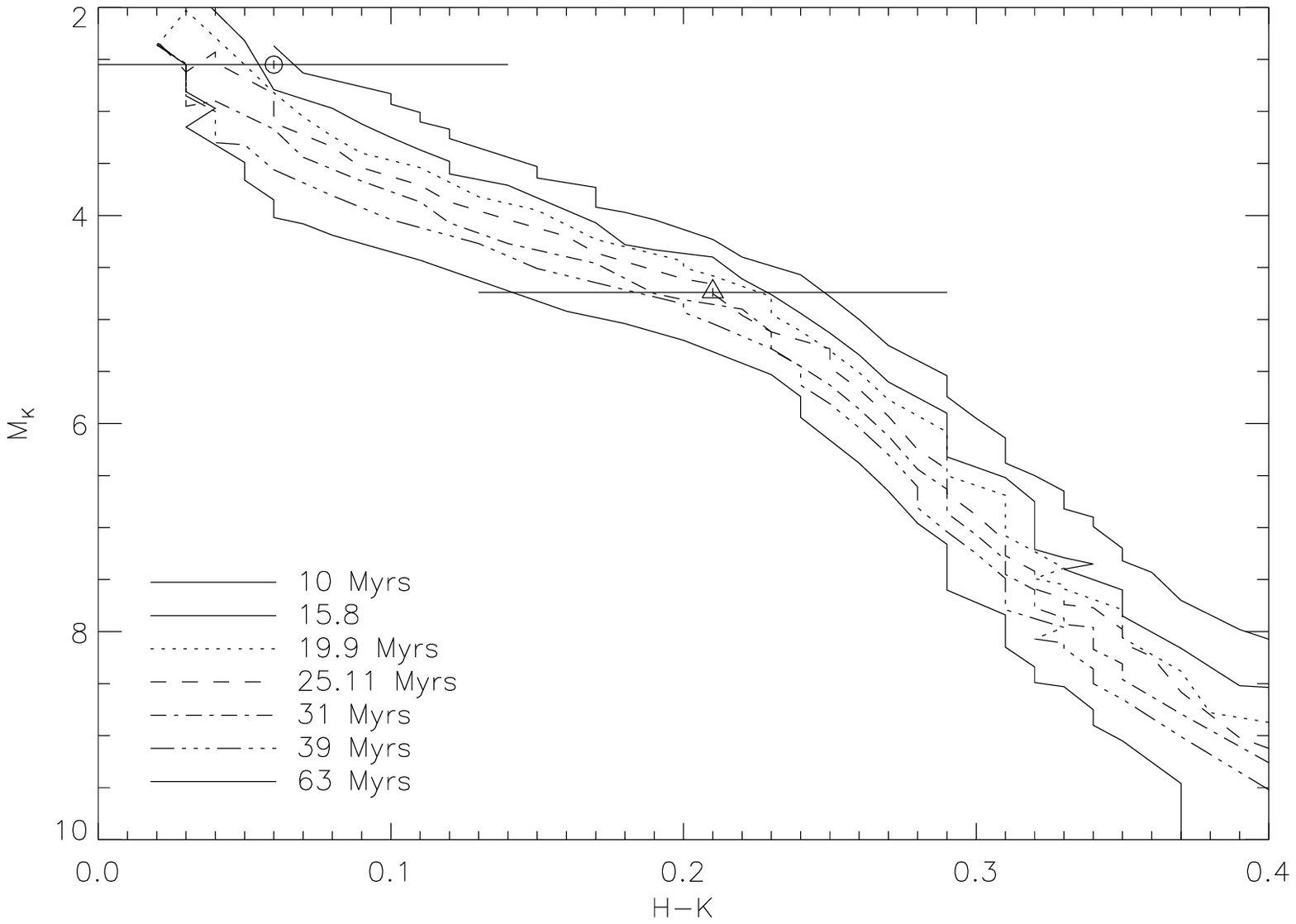}
\includegraphics[width = 7.7cm]{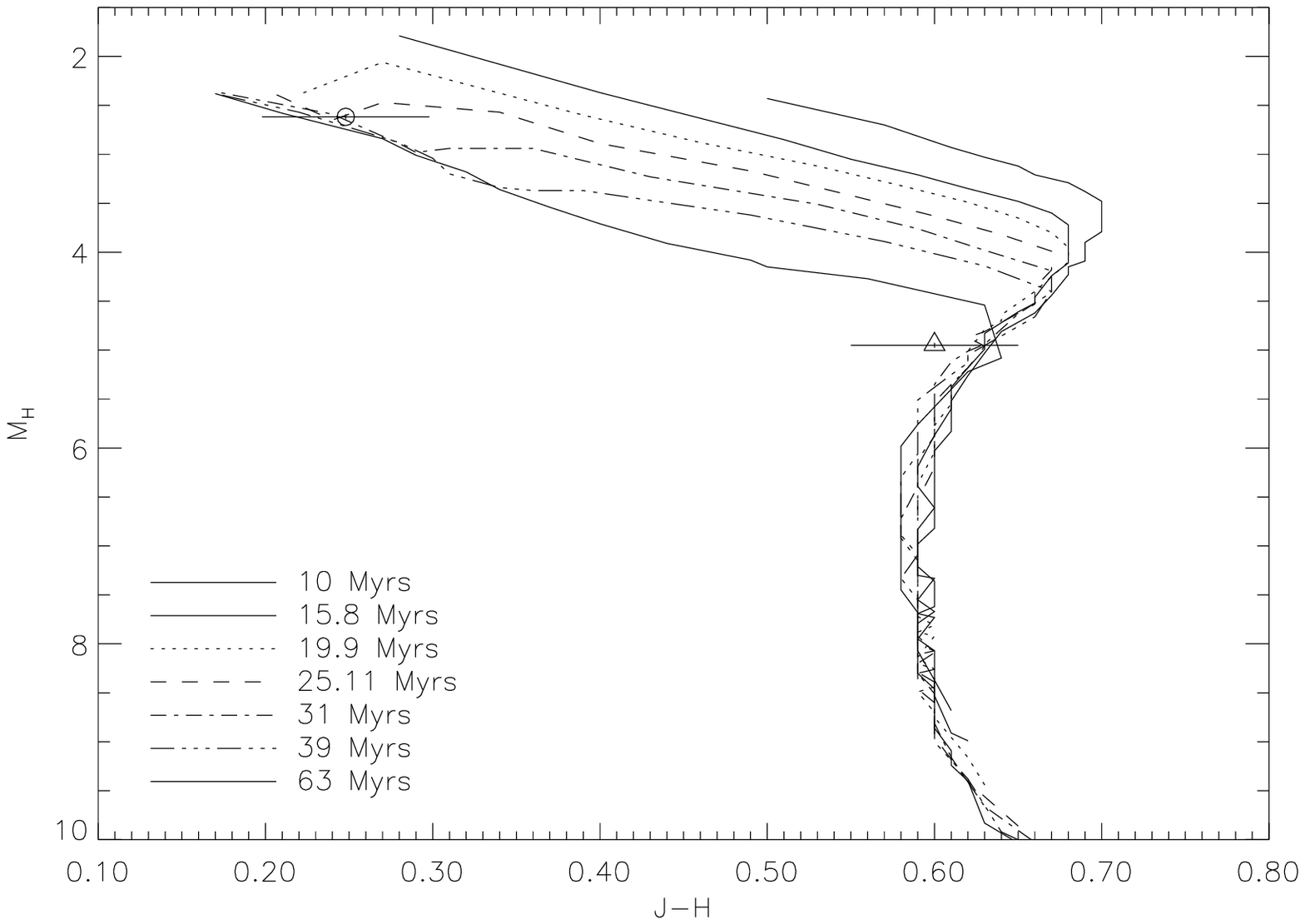}
\caption[]{\label{fig:iso}{\small Comparison of evolutionnary tracks given by Baraffe et al. (2002) representing absolute Magnitudes versus colors with the measurements of HD~199143~A (circle) and HD~199143~B (triangle). Uncertainties on the colors measurements are also overplotted. We supposed that the system was located at 48 pc with no extinction (Van den ancker et al. 2000; Jayawardhana \& Brandeker 2001).}}
\end{center}
\end{figure}

With our new photometric measurements, we have investigated the evolutionary status of HD~199143~A. Fig.~\ref{fig:iso} shows the colors of HD~199143~A overplotted on the isochrones provided by Baraffe et al. (2002). We thus derive the two possible stellar parameters of this star. The results are presented in Table \ref{tab:stell}. 

These parameters are consistent with an expected F8V spectral type (Schmidt-Kaler et al. 1982; Van den Ancker et al. 2000). The age found for HD~199143~A is older than 18~$\pm$~2~Myr when compared to the estimation reported by Jayawardhana \& Brandeker (2001), but in reasonable agreement if we consider the uncertainty of the color measurements and of the stellar model we used. If we consider the evolutionary tracks $M_{K}$ versus $J-K$ and $M_{H}$ versus $J-H$ which most constrain the age of HD~199143~A, we see that this object is likely to be aged of 25~Myr. However, an age of 31~Myr is still possible according to the uncertainties. We then considered both ages to derive the stellar parameters presented Table~\ref{tab:stell}.    

\subsection{HD~199143~B}

From the color analysis proposed by Jayawardhana \& Brandeker (2001), HD~199143~B appeared to be extremely red. They attributed the strong $J-K$ color of 1.37 magnitude to an IR excess caused by a circum-secondary disk around HD~199143~B, as first suggested by van den Ancker et al. (2000) based on 12 $\mu$m IRAS excess. 
Our more recent observations (Table \ref{tab:phot}) fail to support the need for this interpretation as HD~199143~B is significantly less red ($J-K = 0.81$). As we suppose that the components HD~199143~A and B are coeval, the HD~199143~B age is 25 or 31~Myr. Using the evolutionary model of Baraffe et al. (2002), we derived the stellar parameters for HD~199143~B as for HD~199143~A. The results are presented in Table~\ref{tab:stell}.
These values are consistent with the IR colors of a M2 star according to the color table given by Kenyon et al. (1995). The presence of circumstellar matter is no longer needed to explain the color measurements of this object.

\begin{table}[htb]
     \caption[]{Stellar parameters derived from the isochrones provided by Baraffe et al. (2002) for an age of 25~Myr and 31~Myr.}
        \label{tab:stell}
      \[
         \begin{array}{lllll}
            \hline
            \hline
\noalign{\smallskip}
\multicolumn{5}{c}{\mathrm{Age~of~25~Myr}}\\
\noalign{\smallskip}
\hline
            \noalign{\smallskip}
            \mathrm{Source}     & \mathrm{M}              &  \mathrm{T}_{eff}      & \mathrm{log(g)}   &\mathrm{L}             \\
                       & \mathrm{(M}_{\odot}\mathrm{)}    &  \mathrm{(K)}          &     &\mathrm{(L_{\odot})}  \\
        \noalign{\smallskip}
            \hline
        \noalign{\smallskip}
           \mathrm{HD~199143~A} & [1.2, 1.3]       & [5770, 5990]      & [4.17, 4.29] &[2.18, 2.23]     \\
           \mathrm{HD~199143~B} & [0.57, 0.6]      & [3622, 3651]      & [4.46, 4.45] &[0.08, 0.09]     \\
        \noalign{\smallskip}
            \hline
         \end{array}
      \]
      \[
         \begin{array}{lllll}
            \hline
\noalign{\smallskip}
\multicolumn{5}{c}{\mathrm{Age~of~31~Myr}}\\
\noalign{\smallskip}
\hline
            \noalign{\smallskip}
            \mathrm{Source}     & \mathrm{M}              &  \mathrm{T}_{eff}      & \mathrm{log(g)}   &\mathrm{L}             \\
                       & (M_{\odot})    &  (K)          &     &(L_{\odot})  \\
        \noalign{\smallskip}
            \hline
        \noalign{\smallskip}
           \mathrm{HD~199143~A} & 1.3           & 6046           &  4.28 & 2.24          \\
           \mathrm{HD~199143~B} & [0.6, 0.62]      & [3661, 3683]      &  [4.51, 4.50]& [0.08, 0.09]     \\
        \noalign{\smallskip}
            \hline
         \end{array}
      \]
\end{table}

\section{Origin of the 12 $\mu$m excess}

If a circumsecondary disk is no longer needed to explain the near-IR observations of HD~199143~B, one question is still unsolved: what is the source of the 12 $\mu$m excess detected by Van den Ancker et al. (2000) from IRAS data?  In their spectral analysis, 
Van den Ancker et al. (2000) fitted the optical measurements of the HD~199143~A+B by a single Kurucz spectra (T$_{eff}= 6200$K and log g = 4.3). Extrapolating this spectrum to IR wavelengths, they deduced the existence of an IR excess only at 12 $\mu$m of 0.24~$\pm$~0.04~Jy. At 25, 60 and 100~$\mu$m, respective upper limits of 0.12, 0.12 and 0.3~Jy were obtained from IRAS data.

\begin{figure}[h]
\begin{center}
\vspace{0cm}
\hspace{0cm}\includegraphics[width = 8cm]{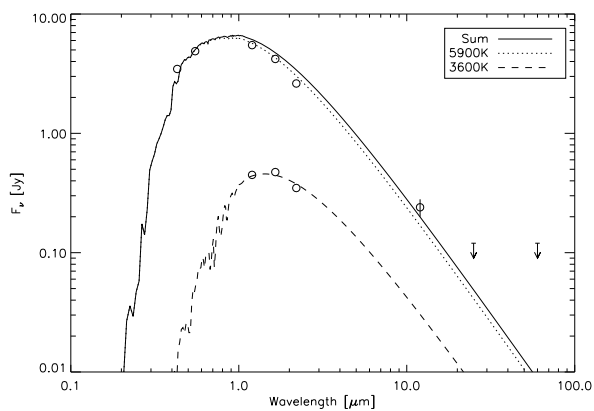}
\caption[]{\label{Fig 3}{\small Observed fluxes of HD~199143 A and B in optical, near-IR and mid-IR (circles) compared to Kurucz spectra of a 5900K star, of a 3600K star and of their sum.}}
\label{fig:12}
\end{center}
\end{figure}

 If we now consider that HD~199143 is a binary system with a F8 primary star and a M2 companion, the spectral contribution of the two stars can be adjusted using optical and near-IR measurements. 
Fig. \ref{fig:12} shows Kurucz (1991) spectra of a 5900K star for HD~199143~A and of a 3600K star for HD~199143~B. Their sum and observed measurements (circles) are also plotted.
If we compare the sum of the two spectral contributions of HD~199143~A and HD~199143~B to the IRAS measurement at 12 $\mu$m, the IR excess is no longer obvious. Indeed, considering the uncertainty on the IRAS flux, the spectral sum of the two components is consistent with the value of 0.24~$\pm$~0.04~Jy. Therefore, the IRAS 12 $\mu$m observation can be explained by a late spectral type for HD~199143~B. Hence, the presence of dust in this system is no longer necessary to explain near-IR and mid-IR measurements.

\section{Conclusions}

We have presented new adaptive optics images of the binary system HD~199143~AB from the Capricornus Association. They confirm the two components of this source, as resolved by Jayawardhana \& Brandeker (2001), but return a different contrast in $J$ band. We discussed the possible causes of this difference (photometric variability, AO correction, deconvolution algorithms). By comparing the results obtained with IDAC and MISTRAL deconvolution techniques, we found that the data are consistent with observations obtained under good AO correction. For less favorable conditions, the regularisation of the PSF and/or the object in the deconvolution process is critical and may be responsible for the discrepancy between the two algorithms on the observations of Jayawardhana \& Brandeker (2001).

Our new results confirm the F8 spectral type of HD~199143~A but derive different characteristics for its companion HD~199143~B. From our observations, we have deduced that HD~199143~B is more likely to be a M2 star. From the evolutionnary model of Baraffe et al. (2002), we have derived the stellar parameters for the two stars and found an estimation for the age of the system between 25~Myr and 31~Myr, slightly higher than Jayawardhana \& Brandeker (2001).
We have also found that the IR excess at 12~$\mu$m is no longer needed because the spectral sum of the components HD~199143~A and HD~199143~B was consistent with the IRAS flux of 0.24~$\pm$~0.04~Jy. Consequently, we conclude that the existence of the late-type companion HD~199143~B can explain the present near-IR and mid-IR observations.

\begin{acknowledgements}
We would like to thank Ray Jayawardhana and Alexis Brandeker who kindly shared their observations. This allowed us to perform tests on both myopic deconvolution algorithms IDAC and MISTRAL and to conduct our discussion on the contrast estimation. We would like to thank also the CNES for the financial support given to Jean-Charles Augereau and finally the staff of the ESO 3.6m telescope and particulary Kate Brooks for her outstanding support.

\end{acknowledgements}



\end{document}